\documentclass[usenatbib]{mn2e}
\usepackage{epsfig}
\usepackage[pass]{geometry}
\usepackage{times}
\usepackage{graphicx}

\title[The continued echo of SN~2002hh]{Late-time spectroscopy of
  SN~2002hh: A continued visible light echo with no shock interaction yet}

\author[Andrews et al.]{J.E. Andrews$^1$\thanks{Email:
    jandrews@as.arizona.edu}, Nathan Smith$^1$, \& Jon C.\ Mauerhan$^2$ \\$^1$Steward
  Observatory, University of Arizona, 933 North Cherry Avenue, Tucson,
  AZ 85721, USA\\$^2$Department of Astronomy, University of California, Berkeley, CA 94720-3411, USA}

\begin{document}
\date{Accepted 0000, Received 0000, in original form 0000}
\pagerange{\pageref{firstpage}--\pageref{lastpage}} \pubyear{2002}
\def\arcdeg{\degr}
\maketitle
\label{firstpage}

\begin{abstract}

  Supernova (SN) 2002hh was unusual among core-collapse SNe because it was highly reddened, and displayed a bright infrared (IR) excess due
  to radiatively heated dust in its circumstellar medium (CSM). Estimates for the mass of dust responsible for the IR echo suggested the presence of a massive shell within 0.26 pc of the star.  For a velocity of 5000 - 10000 km s$^{-1}$, this material should be hit by the SN blast wave at late times, starting at roughly 12 years  post-explosion. We have obtained deep late-time spectra with the MMT Blue Channel spectrograph to search
  for any spectral signatures of ongoing shock interaction. Interaction with a strength comparable to SN 1987A's collision with the equatorial ring would be detected in our data.  However, in the spectra reported here, we do not detect clear signs of strong CSM interaction, contrary to expectations based on the reported radii of the dust shell.  We do, however, detect emission associated with the old SN, and we find that the  broad lines in the spectrum indicate a continuation of an ongoing reflected light echo, which appears similar to the spectrum at peak luminosity for this Type II-P event.

\end{abstract}


\section{INTRODUCTION}

Prior to explosion, the progenitors of core collapse supernovae (CCSNe) can experience significant mass-loss.  Depending on various factors, such as progenitor mass, wind velocity and density, and the time between mass-loss events and explosion, a wide range of luminosities and spectral features can be observed as the SN begins to interact with the surrounding environment (for a comprehensive overview, see \citet{2014ARA&A..52..487S}.  Dense, nearby circumstellar material (CSM) that gets hit by the blast wave shortly after explosion can create Type IIn SNe, named for the narrow features found in their spectra \citep{1990MNRAS.244..269S}. These CSM interactions may be detected from the birth of the SN itself, but in some cases nebulous and/or distant CSM may not be detected until months or years after explosion.

 The onset of the interaction between the SN blast wave and the CSM is usually heralded by intermediate-width (10$^{3}$ km s$^{-1}$) emission lines, occasionally with flat-tops or multiple-peaks.  Additionally, the light pulse from the supernova explosion can be scattered off of dust created through this mass-loss, and may reveal the structure and properties of the surrounding medium as well as snapshots into the evolutionary history of the SN progenitor. The light can be purely scattered off of the dust, creating an optical scattered light echo, or can be absorbed and remitted by the dust grains, creating a thermal IR echo. Light echoes have been seen around many types of transients including SNe \citep[for example]{1991ApJ...366L..73C,2001ApJ...549L.215C,2002ApJ...581L..97S,2008ApJ...677.1060W,2013AJ....146...24V}  and eruptive variables \citep[for example]{2003Natur.422..405B,2012Natur.482..375R,2014ApJ...787L...8P}. For a full review, refer to \citet{2012PASA...29..466R}. 

The Type IIP SN 2002hh was discovered on 2002 October 31 \citep{2002IAUC.8005....1L} just southwest of the nucleus of NGC 6946, and shortly thereafter typed as a young Type II SN showing a highly flattened and reddened continuum \citep{2002IAUC.8007....2F}. Early optical photometric evolution indicated that this appeared to be a rather unremarkable object among SN II-P, with a plateau phase lasting $\sim$ 2-3 months, followed by a linear decline of roughly 0.01 mag d$^{-1}$.  Optical spectra over the first year also showed a normal CCSNe transitioning into the nebular phase, with broad H$\alpha$, [O I] $\lambda\lambda$ 6300, 6364, and [Ca II] $\lambda\lambda$77921, 7323 emission lines becoming prominent  \citep{2006MNRAS.368.1169P}.  It was therefore surprising when optical observations obtained 2-4 years post-explosion revealed no significant fading in the light curve, and spectra almost identical to the pre-nebular phase spectra, including an H$\alpha$ line still retaining a P Cygni profile \citep{2007ApJ...669..525W}.  It was postulated that a scattered light echo off of nearby dust  could explain this lack of late-time fading. Subsequent deep Hubble Space Telescope (HST) imaging in 2006 and 2007 did in fact reveal a spatially resolved scattered light echo from a circumstellar shell located $\sim$10$^{19}$ cm from the SN, and with a surface brightness 300 times brighter than the background \citep{2007ApJ...669..525W, 2012ApJ...744...26O}.

Near- and mid- IR observations of SN 2002hh post-plateau revealed another source of pre-existing CSM material roughly 10$^{17}$ - 10$^{18}$ cm away from the explosion site.  First seen as a K - L$^{\prime}$ excess after day 200 \citep{2006MNRAS.368.1169P}, the IR excess was then confirmed with Spitzer Space Telescope IRAC and MIPS observations around day 600 \citep{2006ApJ...649..332M} and Gemini/Michelle 11.2 $\mu$m imaging soon thereafter \citep{2005ApJ...627L.113B}.  Estimates for the mass of dusty CSM around SN 2002hh responsible for this IR excess range from as much as 0.10-0.15 M$_{\sun}$ \citep{2005ApJ...627L.113B}, to as little as 0.04 M$_{\sun}$ \citep{2006ApJ...649..332M}.  The discrepancy arrises in part from the difficulties of disentangling the flux from a nearby, bright star forming region.

From the optical and IR observations we know that SN 2002hh has both an inner CSM with which the SN ejecta will interact on a relatively short timescale, and an outer CSM, likely a contact discontinuity between the red supergiant (RSG) wind and the ambient ISM, which is responsible for a very strong scattered light echo. Using the average  FWHM SN expansion velocity of $\sim$5600 km s$^{-1}$ measured from the spectra presented here, the ejecta should begin interacting with the dust shell only 6 years post-explosion if the smaller radius of 10$^{17}$cm is adopted. This is similar to the nearby, and well studied SN 1987A which has an equatorial ring with a radius or roughly 0.2 pc \citep[0.\arcsec599--0.\arcsec829]{2002ApJ...572..209S} and showed the first substantial signs of shock interaction in 1997, 10 years post-explosion \citep{1997IAUC.6710....2G,1997IAUC.6665....1P}.  Here we report on new medium-resolution spectra of SN 2002hh taken 10-11 years after explosion, which show no indication that the SN ejecta has begun to strongly interact with the CSM.

\begin{figure*}
\includegraphics[width=5in]{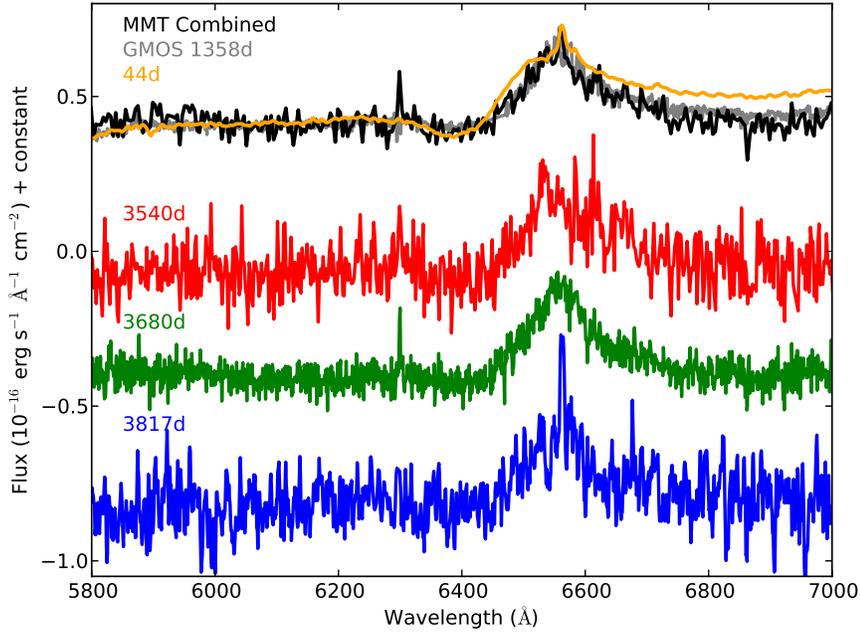}
\caption{ Moderate-resolution MMT spectra from day 3540 (red), 3639/3680 (green), and 3817 (blue).  The summed spectra from all four epochs are shown in black, and overplotted is the Lick day 44 spectra in orange \citep{2006MNRAS.368.1169P} and the Gemini day 1358 spectra in gray \citep{2007ApJ...669..525W}.}
\label{fig:spec}
\end{figure*}

\begin{figure*}
\includegraphics[width=5in]{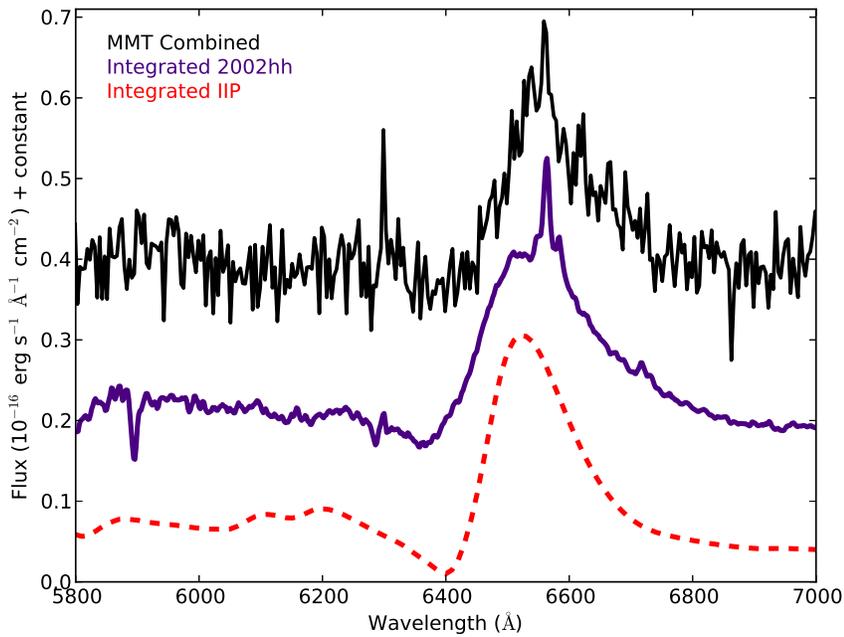}
\caption{The combined MMT spectrum of all epochs compared with a time integrated light-echo spectrum created from early time ($<$ 100 days)  SN 2002hh spectra (indigo line) and model Type IIP spectra (dashed red line).  The similarity in spectrum shape indicates the late time observations of SN 2002hh are in fact dominated by a scattered light echo. }
\label{fig:spec}
\end{figure*}
\begin{figure*}

\includegraphics[width=5in]{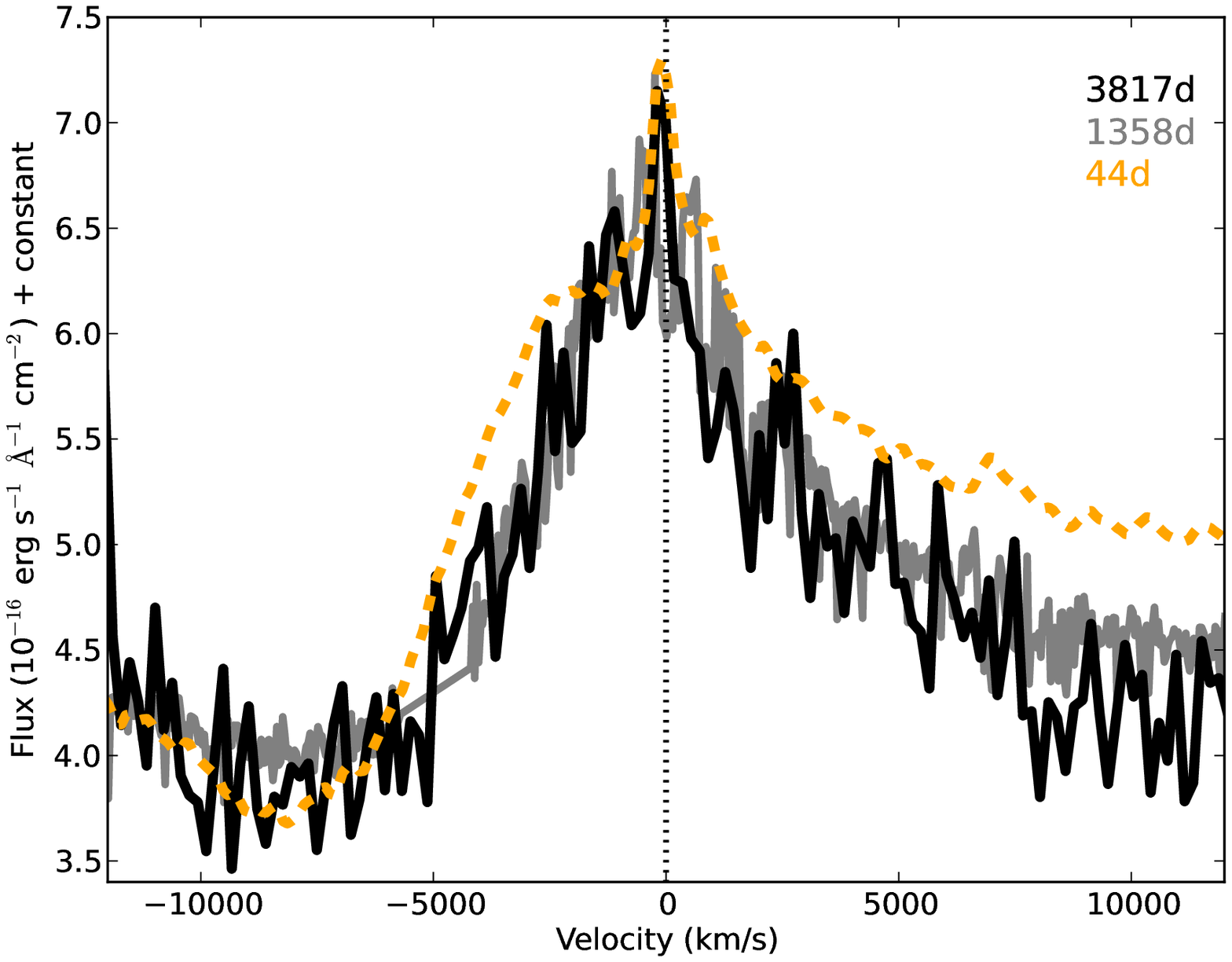}
\caption{The day 44, 1358, and 3817 H$\alpha$ line plotted in velocity-space. For comparison the day 2873 spectrum of the equatorial ring in SN 1987A has been plotted as well.}
\label{fig:spec}
\end{figure*}

\section{OBSERVATIONS}

We obtained 4 epochs of moderate-resolution optical spectra with the Blue Channel spectrograph on the Multiple Mirror Telescope (MMT) on 2012 Jul. 8, Oct. 15, Nov. 25 and on 2013 Apr. 4, roughly 10 years post-explosion (see Table 1).  Each observation was taken with the 1200 l mm$^{-1}$ grating with a central wavelength of 6350 \AA\  and 3 x 1200s exposures. The spectral range covers approximately 5700 - 7000 \AA\ .  Table 1 reports the slit-width used for each observation, as well as the exact spectral range.  Standard reductions were carried out using IRAF \footnote{IRAF, the Image Reduction and Analysis Facility is distributed by the National Optical Astronomy Observatory, which is operated by the Association of Universities for Research in Astronomy (AURA) under cooperative agreement with the National Science Foundation (NSF)}. Flux calibration was achieved using spectrophotometric standards at a similar airmass taken with each science frame.

The final spectra are shown in Figures 1 and 2.  Due to the low signal-to-noise, and apparent lack of temporal evolution, we have combined the 2012 Oct. 15 and Nov. 25 spectra (labeled day 3680).  Additionally we have combined all 4 epochs into one spectrum for comparison with the earlier spectra.  The spectra have been rebinned by a factor of 4 to lower resolution.  A total extinction value of A$_{v}$ $=$ 6.1 (E(B-V)=1.96) was reported by \citet{2002IAUC.8024....1M}, and has been applied to the spectra presented here. We have also applied a redshift correction of z=0.000133 to all spectra shown below. An explosion date of 2002 October 29 \citep{2006MNRAS.368.1169P} is adopted throughout the paper.

\begin{table}
\begin{center}\begin{minipage}{3.3in}
      \caption{Late-Time Optical Spectroscopy of SN~2002hh}\scriptsize
\begin{tabular}{@{}lcccccc}\hline\hline
  Date    &Telescope/Instrument    &Day   &Slit Width & R  & Range     \\ 
  (y-m-d)     &&(max) &    (arcsec)  & $\frac{\lambda}{\Delta \lambda}$ & (\AA)       \\   \hline
2012-07-08 &MMT/Blue Channel  &3540     & 1.0  & 4500  & 5716--7024   \\
2012-10-15 &MMT/Blue Channel &3639     & 1.25   & 4500 & 5696--7003 \\
2012-11-25 &MMT/Blue Channel &3680     & 1.5    & 4500 & 5691--6998   \\
2013-04-11 &MMT/Blue Channel &3817     & 1.0    &  4500 & 5699--7007 \\
\hline
\end{tabular}\label{tab:pcyg}
\end{minipage}\end{center}
\end{table}

\section{RESULTS}
As can be seen in Figure 1, the spectrum of SN 2002hh 10-11 years post-maximum has changed very little since the last published spectrum on day 1353 \citep{2007ApJ...669..525W}. The H$\alpha$ FWHM of 5600 km s$^{-1}$ has persisted, and the P-Cygni shape, prominent in the early-time spectra as well, is still evident, with the red wing appearing broader than the blue. In the day 3540 and 3680 spectra there are narrow lines (FWHM = 85 km s$^{-1}$) of H$\alpha$, [N II], and [S II] that likely originate in a nearby HII region. These lines were also seen by \citet{2006MNRAS.368.1169P} in early spectra of SN 2002hh, and are likely due to a combination of slit width, position, and incomplete background subtraction. The narrow line at 6300 \AA\ though has two possible origins, while it is likely a sky line, it could be [O I] emission from unshocked gas in the surrounding CSM that is being ionized by x-rays from the beginnings of shock interaction. Similar features were seen in SN 1987A \citep{2008A&A...492..481G}, and could be the first indication that interaction has begun.  Due to the low recession velocity of the SN (48 km s$^{-1}$), and the moderate-resolution of the instrument it is difficult to disentangle a terrestrial or extra-galactic origin.

To simulate the potential appearance of a light-echo spectrum, we created a time-integrated spectrum of SN 2002hh, using the early time spectra presented in  \citep{2006MNRAS.368.1169P}  and utilizing the methods presented in  \citep{2012ApJ...749..170S} for the case of SN 1980K. As shown in Figure 2, the results are consistent with the late-time MMT spectrum of SN 2002hh
and thus support the light-echo hypothesis.  Additionally we have created a model light echo spectrum using Type IIP spectral templates made available by Peter Nugent \footnote{https://c3.lbl.gov/nugent/nugent\_templates.html}, and presented in \cite{2004ApJ...616L..91B}.  Other than the deeper P-cygni profile, the general spectrum shape is also very similar to the late-time SN 2002hh spectrum, further strengthening the light-echo argument.

While the Gemini data from \citet{2007ApJ...669..525W} were not flux calibrated, those authors estimate a drop in the H$\alpha$ flux on day 1358 by a factor of 2 from the day 397 spectrum presented in \citet{2006MNRAS.368.1169P}.   We measure an H$\alpha$ flux of 2.0$\times$10$^{-12}$ erg s$^{-1}$ cm$^{-2}$ from the day 397 spectrum, and therefore estimate an H$\alpha$ flux of 1.0$\times$10$^{-12}$ erg s$^{-1}$ cm$^{-2}$ on day 1358. Measurements of the extinction corrected H$\alpha$ flux from the combination of all four MMT spectra presented here yield 4.0$\times$10$^{-13}$ erg s$^{-1}$ cm$^{-2}$.  This means that over the 6 years since the observation on day 1358, the brightness of SN 2002hh has only decreased by about 1 magnitude. This is confirmed from a Gemini-N GMOS acquisition g$^{\prime}$ observation of NGC 6946 taken on 20 June 2010 (GN-2010A-Q-29), or day 2792, from which we measure a V magnitude of 20.96$\pm$0.20.  Day 1273 HST/ACS photometry give V=20.91$\pm$0.11 \citep{2012ApJ...744...26O}, which would suggest that the continuum has not measurably changed in brightness since 2006.

\section{DISCUSSION}

If we assume there is no strong CSM interaction on day 3817, we can offer a lower limit on the inner-radius of the dust shell surrounding SN 2002hh, as well as an upper limit on the progenitor mass-loss rate ($\dot{M}$).  The fastest moving ejecta at 13000 km s$^{-1}$ will have traveled a distance of 0.14 pc and the bulk of the ejecta, moving at $\sim$ 5600 km s$^{-1}$, would have only reached material 1.85$\times$10$^{17}$ cm away.  For comparison, the equatorial ring in SN 1987A has a radius of 0.2 pc, yet the strong CSM interaction began about 10 years after the explosion. The fact that we do not see any signs of ejecta interaction most likely implies that the shell resulting from pre-supernova mass loss is located further from the star than this. For a caveat to this assumption see below. \citet{2006MNRAS.368.1169P} estimate a dust free cavity created from the peak luminosity of the SN to be roughly 1.3$\times$10$^{17}$ cm, similar to the distance the bulk of the ejecta has traveled.

 If we use this distance as a lower limit of the CSM radius we can estimate an upper limit for $\dot{M}$ of the progenitor star.  If we consider a shell mass of 100 M$_{\sun}$ (assuming a normal gas-to-dust ratio of 100), which was the best fit model from \citet{2005ApJ...627L.113B}, and insert these values into equation 19 from \citet{1983ApJ...274..175D} we obtain $\dot{M}$ $<$ 7.0$\times$10$^{-4}$ M$_{\sun}$ year$^{-1}$ if the progenitor wind speed is 10 km s$^{-1}$.  Of course the shell will not be infinitely thin, and an outer radius of 1$\times$10$^{18}$ cm may be more likely. Combining a larger shell radius with a lower estimated shell mass of 4.0 M$_{\sun}$ gives us a lower value of $\dot{M}$ $<$ 1.2$\times$10$^{-4}$ M$_{\sun}$ year$^{-1}$. Additionally the progenitor wind-speed may be higher than the normal RSG winds of 10 km s$^{-1}$ for part or most of the life of the star, which would also alter the mass-loss rate.

  The K - L$^{\prime}$ excess that arose sometime between day 200 and 314 can also yield an estimated CSM inner-radius.  If this excess was caused by an IR echo from the initial SN pulse being absorbed and re-emitted by the dust in the surrounding CSM, then the inner radius of the dusty CSM likely ranges between 5.2 $\times$10$^{17}$ -  8.1 $\times$10$^{17}$ cm.  This is roughly consistent with the current lack of CSM interaction, but it may also suggest that strong CSM interaction is likely to turn on soon. For reference, the radii of the equatorial ring around the blue supergiant SBW1, a twin to pre-explosion SN 1987A, is also roughly 0.2 pc \citep{2007AJ....133.1034S,2013MNRAS.429.1324S}, providing further motivation to expect strong interaction to begin once the ejecta has reached this distance.

It is important here to point out that the scattered light echo resolved in HST images and reported in \citet{2007ApJ...669..525W}, \cite{2012ApJ...744...26O}, and a forthcoming paper by Sugerman et al. (2015) is not produced by the same dust as the 0.04 -- 0.15 M$_{\sun}$ of dust responsible for the IR excess.  The scattered light echo, with an angular radius of 0.$\arcsec$18, is located over 1$\times$10$^{19}$ cm (12 ly) away from the supernova, two orders of magnitude farther than the CSM.  At this distance it will take the ejecta almost 250 years to reach.  This does not mean a brief scattered light echo did not result from the initial flash passing through the closer CSM material, but that would have occurred less than a year after explosion and would have been too near the bright SN to resolve. If SN 1987A was moved to the distance of SN 2002hh (5.9 Mpc), the size of the triple-ring system would only span about 0.$\arcsec$04, which would make it unresolvable from the SN light, even with HST.  

If in fact the inner edge of the CSM is located at a distance of $\sim$ 6$\times$10$^{17}$ cm, we may begin to see signs of interaction in mid- 2015.  As the blast wave from the SN begins to interact with the CSM a two shock system is produced, wherein the forward moving ejecta will crash into the CSM and a reverse shock will plow into the expanding ejecta \citep{1982ApJ...258..790C}.   If the forward shock is radiative, this will manifest itself as narrow to intermediate width emission lines emerging as the shock interaction begins, similar to the 200 km s$^{-1}$ lines seen in SN 1987A \citep{2002ApJ...572..906P}. These will appear in tandem with broad lines, both from the light echo and from the non-radiative reverse shock \citep{2005ApJ...635L..41S}. If we assume a scenario similar to that of SN 1987A, we would also expect the soft X-ray and radio synchrotron emission luminosity  to increase as the forward shock begins to interact with the dense medium \citep{2008A&A...492..481G,2007AIPC..937...86G}. The strength of the intermediate-width emission lines should also increase with time, as more and more gas is swept up.  

Emission line shapes may serve as a diagnostic of the CSM geometry. Multiple intermediate-velocity emission line components, like those seen in SN 1993J \citep{2000AJ....120.1499M}, SN 1998S \citep{2000AJ....119.2968G}, SN 2007od \citep{2010ApJ...715..541A}, and PTF11iqb \citep{2015arXiv150102820S} may also become apparent, if the expanding CSM is not mainly in the plane of the sky. 

As alluded to above, we also have to consider the fact that the interaction may already be occurring, but that the luminosity of the light echo swamps the shock emission. With a scenario similar to that of SN 1987A, we can estimate the relative fluxes of the emission of each of the components, the CSM and the ejecta, as the shock interaction begins.  \citet{2012ApJ...744...26O}, using high resolution imaging of SN 2002hh with HST, estimates that 80$\%$ of the flux at late times is from the light echo. If we assume that this can also be applied to the H$\alpha$ emission, then the ejecta contribution to the H$\alpha$ line is only 8$\times$10$^{-14}$ erg s$^{-1}$  cm$^{-2}$.  We know that the ring emission in SN 1987A was roughly 2 magnitudes brighter on day 3800 than the ejecta emission \citep{2003LNP...598...77L}, which would translate into a shock interaction flux for SN 2002hh of 6.3$\times$ that of the H$\alpha$ emission from the ejecta, or 5$\times$10$^{-13}$ erg s$^{-1}$ cm$^{-2}$, a value consistent with light echo emission.  Therefore if there were interaction in 2002hh, with a strength akin to SN 1987A at a similar epoch, then we should be able to detect it in the optical spectra, particularly in the H$\alpha$ emission. For this reason we conclude the strong CSM interaction with the shell causing the IR echo has not yet begun.

Furthermore, while the CSM interaction of SN 1987A was heralded by the appearance of the first hot-spot at $\sim$10 years, the full illumination of the equatorial ring took over a decade to attain.  The H$\alpha$ flux of the shock interaction 18 years post-explosion was measured to be $\sim$ 2 $\times$ 10$^{-13}$ erg s$^{-1}$ cm$^{-2}$, or over 20 times greater than when the first hot-spot was observed \citep{2005ApJ...635L..41S, 2002ApJ...572..209S}. At the distance of SN 2002hh, this flux would only be 1.5 $\times$ 10$^{-17}$ erg s$^{-1}$ cm$^{-2}$, significantly below the light echo flux. This is not completely surprising, as the peak magnitude of SN 1987A was $\sim$2 mags fainter than that of SN 2002hh \citep{1993AJ....105.1895S}. The HST imaging, as mentioned above, leads us to infer a flux at the start of interaction to be 5 $\times$ 10$^{-13}$ erg s$^{-1}$  cm$^{-2}$, already much brighter than for SN 1987A at a similar phase. Assuming this will only be a portion of the total shock interaction and that the shock evolution will follow that of SN 1987A, we can estimate the shock interaction flux of SN 2002hh will rise to even higher values over the next 8 years. 

If the scattered light echo that is responsible for the heightened emission of SN 2002hh is created by a sphere with a thin shell, the radius of 12 light years is especially important and fortuitous in observing the start of CSM interaction.  In late 2014 and throughout 2015, the far-side of the echo material should be illuminated as the light pulse reaches the shell, and we should begin to see a decline in the light echo brightness as it moves through. This means that it will become even more likely that we will see observational signs of the shock interaction beginning, since the fading echo will have a harder time hiding the shock emission. Continued monitoring of SN 2002hh, particularly over the next few years, is needed to determine the precise time when the ejecta begin to collide with the pre-supernova mass-loss shell.  While it is true that the production of  type IIn SNe require nearby shells ejected just before core collapse, the fraction of SNe with more distant CSM ejected centuries or millennia before the SN is still poorly constrained. Future observations of SN 2002hh will provide tight constraints on the distance and physical properties of the surrounding CSM, and can provide invaluable information into the mass loss history of massive stars and into the evolution of CCSNe.

\smallskip\smallskip\smallskip\smallskip
\noindent {\bf ACKNOWLEDGMENTS}
\smallskip
\footnotesize

We would like to thank the referee for valuable suggestions, as well B.E.K. Sugerman for private communications involving proprietary HST data as well as integrated light echo modeling.  Some observations reported here were obtained at the MMT Observatory, a joint facility of the University of Arizona and the Smithsonian Institution. N.S. received partial support from NSF grants AST-1210599 and AST-131221.


\bibliographystyle{mn2e}
\bibliography{2002hhbib}

\end{document}